\documentclass[sn-basic,Numbered]{sn-jnl}
\usepackage[utf8]{inputenc}
\usepackage{etoolbox}
\usepackage{graphicx}
\graphicspath{{assets/}{assets_compl/}}
\usepackage{mathtools}
\usepackage{amsmath}

\usepackage{amssymb}
\usepackage[overload]{empheq}
\usepackage{siunitx}
\usepackage{longtable,tabularx}
\usepackage{multirow}
\usepackage[capitalise]{cleveref}
\usepackage{subfigure}
\usepackage[export]{adjustbox}
\usepackage{booktabs}
\usepackage{bookmark}
\usepackage{stfloats}
\usepackage{tikz}
\DeclareRobustCommand{\tikzline}[1]{\raisebox{2pt}{\tikz{\draw[black,solid,line width=0.9pt,#1] (0,0) -- (5mm,0);}}}
\usepackage{bm}
\usepackage{bbold}
\newcommand{\mat}[1]{ \bm{#1} }
\renewcommand{\vec}[1]{ \bm{#1} }

\begin{document}
\title{Taming the Tilt: A Unified Pilot Control Concept for Transformational eVTOL Aircraft}
\author*[1,2]{\fnm{Daniel} \sur{Milz} \orcid{0000-0001-9704-2036}}\email{daniel.milz@dlr.de}
\author[1,2]{\fnm{Marc} \sur{May} \orcid{0009-0007-1309-9811}}\email{\{firstname\}.\{lastname\}@dlr.de}
\author[1]{\fnm{Andreas} \sur{Seefried} \orcid{0000-0002-8367-2704}}
\author[1]{\fnm{Tobias} \sur{Bellmann} \orcid{0000-0002-5897-6191}}
\affil*[1]{\orgdiv{Institute of Flight Systems}, \orgname{German Aerospace Center (DLR)}, \orgaddress{\street{M\"unchner Str. 20}, \city{We\ss{}ling}, \postcode{82234}, \country{Germany}}}
\affil[2]{\orgdiv{TUM School of Engineering and Design}, \orgname{Technical University of Munich (TUM)}, \orgaddress{\street{Lise-Meitner-Str. 9}, \city{Ottobrunn}, \postcode{85521}, \country{Germany}}}
\abstract{
Transformational electric vertical take-off and landing (eVTOL) vehicles have gained significant attention over the past decade due to their efficient wing-borne cruise capabilities and reduced reliance on ground-based infrastructure. However, control system design for these vehicles remains challenging, as they must operate across multiple flight phases, each with distinct dominant dynamics. If left unaddressed, this complexity would significantly increase pilot workload, thus motivating the development of pilot control systems for multi-phase flight operations. The Simplified Vehicle Operations concept presents a promising strategy for reducing pilot workload. This study presents the design and implementation of a novel pilot control concept for eVTOL aircraft, validated through a tandem tilt-wing aircraft simulation on a full-motion simulator equipped with an active, force-feedback side stick. The system provides pilots with tactile feedback during specific flight phases, supporting intuitive control. The proposed approach enables seamless transitions and multi-phase flight maneuvers by leveraging the available degrees of freedom. 
Furthermore, an optimal-control-based methodology is proposed as a metric to evaluate command-filter-induced performance penalties and inceptor activities. The results show that the proposed command filter does not significantly increase the mission duration compared to the closed-loop system, while the active side stick helps reduce inceptor activity.}
\keywords{Flight Control, Flight Simulation, Pilot Control Concepts, Motion Simulation, Optimal Control, Human-Machine Interface}
\maketitle
\section*{Notation convention}
Throughout this paper, vectors are written in small bold letters (e.g., $\vec{v}$), matrices in bold capital letters (e.g., $\bm{J}$, $\mathbf{R}_{NB}$), and scalars in italic (e.g., $m$, $t$). Time derivatives of states are written in differential form (e.g., $\frac{d\vec{x}}{dt}$), while commanded or measured rates are denoted with dot notation (e.g., $\dot{h}_{\mathrm{com}}$). The nomenclature and symbol definition are shown in~\cref{tab:nomenclature}. The coordinate frame of a vector is denoted as a superscript and can be the inertial frame N (North-East Down, NED), B (body), C (control, cf.~\eqref{eq:vc}), or K (kinematic). The components of a vector, e.g., \( \mathrm{x} \), \( \mathrm{y} \), and \( \mathrm{z} \), are denoted as subscripts.
\begin{table}[!htb]
    \centering
    \caption{Nomenclature and symbols}
    \label{tab:nomenclature}
    \begin{tabular}{ll}
        \toprule
        Symbol & Description \\
        \midrule
        \multicolumn{2}{c}{General Flight Dynamics and States} \\
        \midrule
        \( V_\mathrm{gnd}, V_\mathrm{cas} \) & Velocity over ground/calculated airspeed \\
        \( h \) & Altitude \\
        \( \gamma \) & Flight path angle \\
        \( \chi \) & Flight path course \\
        \( \alpha \) & Angle of attack \\
        \( \beta \) & Sideslip angle \\
        $\vec{r}^{N}$ & Position vector in NED frame \\
        $\vec{\theta} = [\phi,\theta,\psi]^T$ & Euler-angle vector (roll, pitch, yaw) from NED to body frame \\
        $\vec{v}^B $ & Velocity vector in body frame w.r.t. inertial frame \\
        $\vec{\omega}^{B}$ & Angular rate vector in body frame w.r.t. inertial frame \\
        $\vec{f}^{B}$ & Force vector in body frame \\
        $\vec{m}^{B}$ & Moment vector in body frame \\
        $\bm{J}$ & Inertia matrix \\
        $m$ & Vehicle mass \\
        $g$ & Gravitational acceleration of \SI{9.81}{\meter\per\second} \\
        $\mat{R}$ & Rotation matrices \\
        $\vec{n}^{B}$ & Load-factor vector in body frame~\eqref{eq:loadfactor} \\
        \( h_\mathrm{HAT} \) & Height above terrain \\
        \( T \) & Propeller thrust \\
        \( \delta_\mathrm{w} \) & Tilt wing angle \\
        \( \delta_\mathrm{e} \) & Elevon deflection \\
        \( \delta_\mathrm{ab} \) & Air brake deflection \\
        \( \bar{q} \) & Dynamic pressure \\
        \( S_\mathrm{ab} \) & Reference area of air brake \\
        \( C_\mathrm{W,ab} \) & Drag coefficient of air brake \\
        \( V_\mathrm{min},  V_\mathrm{max} \) & Minimum/maximum velocity \\
        \( V_\mathrm{h/t},  V_\mathrm{t/c} \) & Hover/transition and transition/cruise boundary \\
        \midrule
        \multicolumn{2}{c}{Control System \& Pilot Control Interface} \\
        \midrule
        \( \sigma \) & generic signal \\
        \( \omega_0, \zeta \) & Natural frequency and damping \\
        \( K \) & Gain or parameter \\
        \( r, r_\mathrm{th} \) & Generic reference signal and threshold \\
        \( \delta_\mathrm{x},\delta_\mathrm{y},\delta_\mathrm{z},\delta_\mathrm{h} \) & Stick deflections lon., lat., yaw, and heave \\
        \( \delta_0, R\) & Zero point and resistance of active stick \\
        \( J \) & Cost function to be minimized \\
        \bottomrule
    \end{tabular}
\end{table}
\section{Introduction}
Electric vertical take-off and landing (eVTOL) aircraft designs are emerging within the Low-Altitude Economy and Advanced Air Mobility (AAM) contexts.
Transformational eVTOL aircraft have gained special attention over the past decade due to their efficient wing-borne cruise capabilities and reduced reliance on ground-based infrastructure.
However, these aircraft present distinct control challenges due to their operation across multiple flight regimes. Transformational VTOLs, in particular, must transition seamlessly between helicopter-like thrust-borne and airplane-like wing-borne flight. This transition involves complex aeropropulsive effects that influence the aircraft’s stability, control authority, and responsiveness to pilot inputs~\cite{Simmons2022,Droandi2018}. 
Early VTOL prototypes, such as the Bell XV-3 and XV-15~\cite{Thomason1990}, required significant pilot skill and manual coordination of control surfaces, engine power, and pylon tilt angles, which contributed to high cognitive demands and increased accident rates~\cite{Maisel2000}.
The Harrier jet encountered a similar issue, referred to as the ``three-hand problem,'' where pilots needed to operate the thrust lever, stick, and nozzle angle lever concurrently during takeoff and landing~\cite{nordeen2006,Lombaerts2020b}.
The adoption of digital fly-by-wire (FBW) systems has significantly simplified pilot control. FBW technology decouples pilot inceptor inputs from direct control surface actuation, enabling the flight control system to interpret pilot commands and coordinate appropriate outputs. Advanced control functions, including Rate Command/Attitude Hold (RCAH) and Translational Rate Command (TRC), further reduce pilot workload and enhance artificial stability~\cite{Lombaerts2020b}.
The transition to FBW systems, e.g. in the Bell Boeing V-22 Osprey~\cite{McManus1985}, abstracts underlying flight physics from the pilot, providing a consistent control interface across all flight regimes and improving safety and operational effectiveness~\cite{Robinson1989}.
Successors to the AV-8B Harrier, including the Vector thrust Aircraft Advanced Control (VAAC) Harrier~\cite{nordeen2006} and the X-35 program~\cite{Walker2002}, have advanced unified control concepts.
The F-35B employs a unified flight control scheme with active pilot inceptors, such as an active sidestick, an active lever, and yaw pedals~\cite{Wurth2018,Harris2018,denham2008}.
Contemporary VTOL designs incorporate advanced control architectures, extending traditional FBW capabilities with control strategies that address transition-flight complexity. As a result, different control functions can be integrated allowing precise maneuvering, improved handling qualities, and a unified control philosophy. There, pilots use the same control commands throughout all flight phases, thus reducing mode confusion~\cite{Angelov2021}. Furthermore, active side sticks have gained increased attention~\cite{Altunkaya2024} as they provide the pilot with tactile feedback, supporting their situational awareness, leading to a more intuitive control.
Although multiple control concepts for VTOL aircraft have been introduced recently, they are all characterized by a complex implementation due to the multi-mode approach and explicit switching between the modes~\cite{Lombaerts2020b,Angelov2021,Dollinger2021}. A continuous switching or blending approach was introduced to tackle these shortcomings in~\cite{Angelov2022}. Furthermore, most concepts use a passive dual-inceptor layout to allow control over multiple degrees of freedom (coming from an attitude control scheme).
In this paper, \emph{unified control} denotes a consistent command semantics across all flight phases, while the inner-loop flight control and allocation layers continuously manage configuration-dependent actuation. The pilot therefore neither needs to select or switch modes nor consider mode-specific properties.
While unified control schemes address transition and input coordination challenges, the next evolution is Simplified Vehicle Operations (SVO), which is especially relevant for AAM applications~\cite{Wing2020,Lombaerts2020b}. The aim is to enable safer operation by less-experienced pilots through lower workload and reduced training demand~\cite{Wing2020}. 
SVO relies on task-oriented, decoupled flight control, where pilots can directly command desired trajectories rather than manage aircraft attitude. Key features include flight-path-centered controls that govern the aircraft’s velocity vector and robust flight envelope protection to maintain safe operational boundaries~\cite{Wing2020}. 
Current AAM control-concept research increasingly targets unified, task-level command interfaces that enable single- or dual-inceptor management of complex multi-phase maneuvers and reduce mode confusion~\cite{Angelov2021,Dollinger2021,Kaneshige2024}. NASA’s Simplified Vehicle Control (SVC) concept further enhances eVTOL operations through vector-based commands and ``hover buttons''~\cite{Kaneshige2024}.
This holistic design approach enhances safety and operational feasibility, broadening access to VTOL flight for future applications.
\begin{table*}[!hbt]
    \centering
    \caption{Possible assignments of inceptors and control inputs during the different flight phases. The first two concepts can be operated with dual inceptors by assigning the hat-switch functionality to the left stick. The last concept blends the variables throughout the transition phase (\( \rightarrow \)).}\label{tab:inceptor_assignments}
    \begin{tabular}{c|r|ccc}
        \toprule
        & & Hover & Transition & Cruise \\
        \midrule
        \multirow{4}{3cm}{Single Inceptor Unified Control~\cite{Lombaerts2020b}} & Stick lon. & \( \dot{h} \) & \( \gamma \) or \( \dot{h} \) & \( \gamma \) or \( \dot{h} \) \\
        & Stick lat. & \( v^C_y \) & \( \phi \) & - \\
        & Stick yaw & \( \dot{\psi} \) & \( \beta \) & \( \beta \) \\
        & Hat lon. & \( \dot{v}^C_x \) & \( V_\mathrm{gnd} \) & \( V_\mathrm{cas}\) \\
        \midrule
        \multirow{4}{3cm}{Single Inceptor E-Z-Fly~\cite{Lombaerts2020b}} & Stick lon. & \( \dot{h} \) & \( \dot{h} \) & \( \dot{h} \) \\
        & Stick lat. &  \( \dot{\psi} \) &  \( \dot{\psi} \) &  \( \dot{\psi} \) \\
        & Hat lon. & \( \dot{v}^C_x \) & \( \dot{V}_\mathrm{gnd} \) & \( \dot{V}_\mathrm{cas}\) \\
        & Hat lat. & \( v^C_y \) & \( v^C_y \) & \( v^C_y \) \\
        \midrule
        \multirow{4}{3cm}{Dual Inceptor SVO~\cite{Angelov2021,Dollinger2021}} & Left stick lon. & \( \vec{v}^C_x \) & \( V_\mathrm{cas} + \alpha^K \) & \( V_\mathrm{cas} \) \\
        & Left stick lat. & \( \vec{v}^C_y \) & \( \beta^K \) & \( \vec{n}^B_y \) \\
        & Right stick lon. & \( \dot{h} \) & \( \dot{h} \) & \( \dot{h} \) \\
        & Right stick lat. & \( \dot{\psi} \) & \( \dot{\psi} \) & \( \dot{\psi} \) \\
        \midrule
        \multirow{4}{3cm}{F-35B~\cite{Walker2013}} & Left lever & \( \dot{V} \) & \( \dot{V} \) & \( \dot{V} \) \\
        & Right stick lon. & \( \dot{h} \) & \( \rightarrow \) & \( \dot{\gamma} \) \\
        & Right stick lat. & \( \phi \) & \( \rightarrow \) & \( \dot{\phi} \) \\
        & Yaw Pedals & \( \dot{\psi} \) & \( \rightarrow \) & \( \beta \) \\
        \bottomrule
    \end{tabular}
\end{table*}
Lombaerts et al.\ proposed a unified control system for eVTOL vehicles using SVO as these aspects are ``inseparably interconnected''~\cite{Lombaerts2020b}. 
Possible inceptor constellations include a side stick, pedals, and a lever~\cite{Walker2013}; a single (3-axis) side stick with a hat switch~\cite{Lombaerts2020b}; or dual side sticks~\cite{Angelov2021,Dollinger2021}. Relevant control concept mappings are shown in \cref{tab:inceptor_assignments}.
\subsection{Regulatory and Design Requirements}\label{sec:req}
Both FAA and EASA are currently adopting specifications and means of compliance (MOC) for (e)VTOL aircraft~\cite{FAA2024PoweredLiftRule,EASAMOCSCVTOLIssue2}.
The EASA has drafted the Special Condition VTOL 01 (SC-VTOL-01). Relevant for the implementation of flight control systems and pilot control interfaces are mainly the sections \textit{VTOL.2135 Controllability}, \textit{VTOL.2140 Control forces}, \textit{VTOL.2145 Flying qualities}, and \textit{VTOL.2300 Flight control systems}, which specify a Flight Guidance System (FGS).
Those requirements led to our design philosophy for the presented control concept, following these concepts:
\begin{enumerate}
    \item Easy to operate and learn, with a special focus on non-pilots.
    \item Absence of input results in the continuation of the current flight state; the side stick can be released at any time without causing instabilities or critical maneuvers.
    \item Unified control throughout the whole envelope.
    \item Transparent inceptor effect, ensuring that the pilot is always aware of the consequences of each input.
    \item Little performance penalty through the command systems
\end{enumerate}
Based on these requirements, the next subsection introduces the proposed architecture, where the design decisions are influenced by the above criteria.
\subsection{Proposed Control Concept}
This work proposes a novel unified pilot control concept for transformational eVTOL configurations following the requirements 1, 3, and 4~\cref{sec:req}, that
(i) is simple and easy to implement,
(ii) does not involve any direct mode switching,
(iii) uses a single inceptor to control the aircraft throughout the entire envelope and all flight phases,
(iv) does not alter the interpretation of an input channel,
and (v) uses an active side stick to reduce cognitive load, and guide and inform the pilot through all maneuvers.
The concept is demonstrated and evaluated using a tandem tilt-wing eVTOL (\cref{sec:config}) on the DLR PAVSIM motion simulator.
Furthermore, we apply optimal-control-based trajectory optimization from~\cite{May2025jgcd} to estimate the performance penalty imposed by the command filters, to verify the effect of the active side stick, and to investigate the smoothness of optimal inceptor inputs.
The contribution is the integrated combination of unified command filtering, full-envelope inversion-based control, and active side stick force feedback in one operational framework. Furthermore, we propose employing full motion simulation and optimal-control-based analysis to rate and clear pilot control concepts.
The work builds on the flight dynamic models and flight control laws developed in~\cite{May2023,Milz2026jgcd}, and continues the work done in~\cite{Milz2022}.
The selection of controlled variables and the control structure is done \textit{holistically} for the complete envelope, i.e., a \emph{unified inversion}~\cite{Milz2022}.
A simplified but representative full-envelope dynamical model of the tandem tilt-wing comprising the main effects is used and adopted for the use in a real-time simulation framework. \cref{sec:fdm} introduces the flight dynamics and specifies the required flight control interfaces. On this basis, suitable pilot control functions and interfaces are proposed in \cref{sec:pilotcontrol}. The implementation uses command filters, which allow a smooth multi-phase eVTOL flight. An active 3-axis side stick is used. The force-feedback guides the pilot during specific flight phases using tactile feedback.
\section{Flight Dynamics Model and Flight Control Laws}\label{sec:fdm}
The flight dynamics model of a transformational eVTOL (e.g., the tandem tilt-wing aircraft in~\cref{sec:config}) are governed by the nonlinear 6-DoF equations of motion and are characterized by the propulsion system and aero-propulsive forces and moments.
The nonlinear state-space representation of the tandem tilt-wing eVTOL can be described by
\begin{subequations}\label{eq:sys6dof}
	\begin{align}
		\frac{\rm d}{{\rm d} t} {\vec{r}}^N \ & = \mat{R}_\mathrm{NB}(\vec{\theta}) \ \vec{v}^B \\
		\frac{\rm d}{{\rm d} t} {\vec{\theta}} \ \ \ & = \mat{R}_\mathrm{\Phi B} (\vec{\theta}) \ \vec{\omega}^B \\
		\frac{\rm d}{{\rm d} t} {\vec{v}}^B \ & = - \vec{\omega}^B \times \vec{v}^B + \mat{R}_\mathrm{BN}(\vec{\theta}) \vec{g} + \frac{1}{m} \vec{f}^B \label{eq:sys6dof_v} \\
		\frac{\rm d}{{\rm d} t} {\vec{\omega}}^B & = \bm{J}^{-1} \left( - \vec{\omega}^B \times \bm{J} \vec{\omega}^B \right) + \bm{J}^{-1} \vec{m}^B\label{eq:sys6dof_omega}
	\end{align}
with the configuration-dependent input vector \( u \) and the state vector
\( x = \left[ \vec{r}^N ,\  \vec{\theta} ,\ \vec{v}^B ,\ \vec{\omega}^B \right]^T \).
Additionally, the load factor \( \vec{n}^B \) and the velocity in the control frame \( \vec{v}^C \) are added as sensor outputs:
\begin{align}
    \vec{n}^B &= \frac{1}{g} \left[ \frac{{\rm d}\, \vec{v}^B}{{\rm d} t} - \vec{g}^B \right] \label{eq:loadfactor} \\
    \vec{v}^C &= \begin{bmatrix}
        \cos\theta & \sin\phi \sin\theta & \cos\phi \sin\theta \\
        0 & \cos\phi & -\sin\phi  \\
        -\sin\theta & \sin\phi \cos\theta & \cos\phi \cos\theta
    \end{bmatrix} \vec{v}^B \label{eq:vc}
\end{align}
\end{subequations}
Actuator dynamics are modeled following~\cite{Milz2026scitech}. However, to clearly isolate and validate the core pilot control concept, sensor latency, wind and turbulence, as well as failure and other off-nominal conditions are not considered in this work.
\begin{figure*}[!bh]
	\centering
	\includegraphics[]{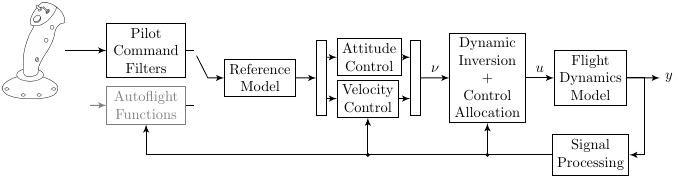}
	\caption{Proposed control architecture for dynamic inversion-based tandem tilt-wing control~\cite{Milz2026jgcd}.}\label{fig:cascade}
\end{figure*}
The overall control architecture is shown in \cref{fig:cascade} and follows~\cite{Milz2026jgcd}. It combines hybrid nonlinear dynamic inversion with optimization-based control allocation. A basic analysis of the applied control law is carried out in~\cite{Milz2026scitech}.
In contrast to similar approaches~\cite{Raab2018,DiFrancesco2016,Panish2023a,Lombaerts2020a,Liu2018}, this work simultaneously inverts the angular rate and velocity dynamics to generate thrust, control surface and tilt angle commands and controls the tilt-angle directly via feedback. 
However, coordination between the attitude and flight path control loops is crucial.
The main focus of this work is on the \textit{pilot command filtering}, which defines behavior across flight regimes. The next section presents these filters in detail.
Dynamic inversion-based architectures are well suited for this purpose as they allow modular function development helping the implementation of requirements in the command filters and reference models. See~\cite{Steinert2025a,Steinert2025b} for a more detailed discussion about dynamic inversion-based control.
\subsection{Flight Control Interface}\label{sec:fcinterface}
An important decision during the control design is the selection of controlled variables. These oftentimes determine the flight control law interface. However, this selection is in general a result of higher-level design requirements.
In our case, both the selection of controlled variables and the design of the pilot control concept is interlaced.
However, the selection needs to ensure adequate control over the rotational and translational motion. While for other aircraft categories, rotation and translation are often cascaded, transformational eVTOLs frequently allow uncoupling both motions. Especially the tandem tilt-wing configuration (\cref{sec:config}) is predestinated for this decoupling.
Thus, following similar works~\cite{Raab2018}, we use the velocity vector in the control frame \( \vec{v}^C \) defined in~\eqref{eq:vc} and the Euler angle vector \( \vec{\theta} \) as a general interface allowing for easy hover and cruise control. While SVO oftentimes makes direct attitude control obsolete, some applications benefit from them. Still, the main command is the translational command.
\subsection{Tandem Tilt-Wing Aircraft Configuration}\label{sec:config}
\begin{figure}[!htb]
	\centering
	    \includegraphics[]{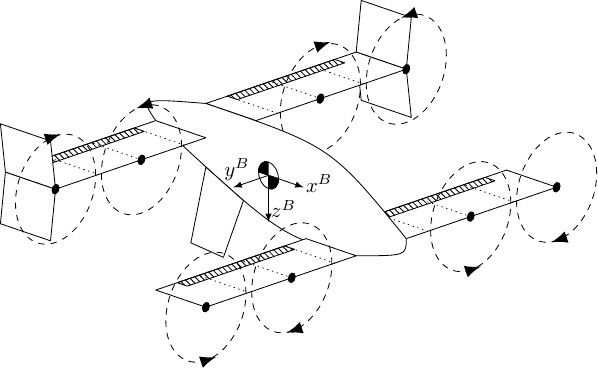}
	\caption{3D sketch of the tandem tilt-wing configuration with annotated input variables.}\label{fig:vahana3d}
\end{figure}

\begin{figure*}[!htb]
	\centering
    \subfigure[Climbing transition maneuver\label{fig:pitch_outboundtransition}]{\includegraphics[page=1, width=.32\linewidth,trim=0 4cm 2cm 0cm,clip]{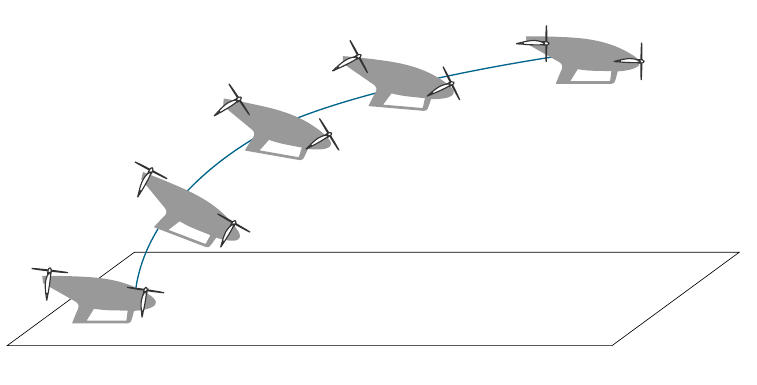}}
	\subfigure[Horizontal motion in hover\label{fig:pitch_hover}]{\includegraphics[page=2,width=.32\linewidth,trim=0 4cm 2cm 0cm,clip]{vahana_transition_slowfast.pdf}}
    \subfigure[Vertical motion during cruise\label{fig:pitch_cruise}]{\includegraphics[page=3,width=.32\linewidth,trim=0 4cm 2cm 0cm,clip]{vahana_transition_slowfast.pdf}}
	\caption{Sketch of the pitch motion supporting the transition governed by slow wing tilt dynamics.}
	\label{fig:pitch}
\end{figure*}

This work uses the tandem tilt-wing flight dynamics model and flight control laws from previous publications~\cite{Milz2022,May2023,Milz2026jgcd}. The aircraft configuration is shown in \Cref{fig:vahana3d} and described in more detail in~\cite{May2023,May2026scitech,Milz2026scitech}.
The configuration is characterized by 14 control inputs and distinguishes itself from other configurations by combining eight electrically driven propellers producing thrust \( T_i \), two independent tandem tilt-wings with angles \( \delta_{\mathrm{w},i} \), and four (one per half-wing) elevons with deflections \( \delta_{\mathrm{e},i} \). The resulting control input vector is
\[ u = \left[ T_{1 \ldots 8}, \delta_{\mathrm{w},1 \ldots 2}, \delta_{\mathrm{e},1 \ldots 4} \right]^T \]
These inputs allow direct control over \( \vec{m}^B \) and over the x- and z-component of \( \vec{f}^B \).
The wings are vertically and longitudinally displaced to reduce interaction effects, especially propeller-propeller and tandem interference. The elevons lie in the wetted surface area of both propellers, which leads to an additional slipstream-interaction effect. Propeller rotation directions are selected to permit nominal-moment cancellation and differential-thrust yaw control in hover.
Compared with prior high-fidelity models, we use a reduced-order model to support real-time simulation and robust piloted testing. The reduced model retains the dominant nonlinear effects, including smooth post-stall aerodynamics and slipstream interactions, through a \textit{minimal strip model} (cf.~\cref{fig:vahana3d}) following \cite{May2023,Cook2021,May2025jgcd,Milz2026scitech}. This approach effectively accounts for the dominant aero-propulsive effects of the tilt-wing dynamics~\cite{May2025jgcd,Simmons2022,Cook2021}.
Furthermore, we assume that both tilt-wing angles are essentially aligned, which is a reasonable approximation~\cite{May2025jgcd}.
To improve handling qualities in the simulator, we additionally introduce air brakes. The primary purpose is to support leveled deceleration during transition, which is a known critical phase in tilt-wing operation~\cite{May2025jgcd}.
The air brakes introduce an additional force that is modeled as pure drag from a surface in free flow, parameterized by the deflection angle \( \delta_\mathrm{ab} \)
\begin{equation}
    \vec{f}_{\mathrm{airbreak},z}^B = \bar{q} S C_W \sin\delta_\mathrm{ab} 
\end{equation}
As illustrated in \cref{fig:pitch} and explained in~\cite{Milz2026jgcd}, pitch attitude control can enhance the performance of tilt-wing longitudinal flight path control.
This can be achieved through a strategy called ``pitch-supported tilting'' similar to Pseudo Control Hedging (PCH)~\cite{Johnson2000}.
Tilt actuators typically have limited bandwidth and therefore constrain transition dynamics. In contrast, pitch response is faster while affecting similar longitudinal dynamics. We therefore use pitch motion to compensate for tilt-angle tracking errors. These errors primarily originate from actuator bandwidth limits and the tilt-angle range constraint of \SIrange{0}{90}{\degree}.
Transformational eVTOL aircraft generally operate in three flight regimes, i.e., hover flight, cruise flight, and the transition between them. For the tandem tilt-wing configuration used here, the phases are defined according to~\cite{Milz2026scitech} as
\begin{itemize}
    \item Hover flight: \( V_\mathrm{min}= \SI{-5}{\meter\per\second} \leq \vec{v}^C_\mathrm{x} \leq \SI{10}{\meter\per\second} = V_\mathrm{h/t} \)
    \item Cruise flight: \( V_\mathrm{t/c} = \SI{37}{\meter\per\second} \leq \vec{v}^C_\mathrm{x} \leq \SI{70}{\meter\per\second} = V_\mathrm{max} \) and \( \delta_\mathrm{w} \leq \SI{12}{\degree} \)
    \item Transition: In between hover and cruise flight
\end{itemize}
\subsection{Reference Models}\label{sec:refmodel}
The reference signals in~\cref{fig:cascade} are generated by a dedicated reference model, which dictates the closed-loop dynamical behavior. This model takes the desired signal \( \sigma_\mathrm{d} \), and optionally its derivative \( \dot{\sigma}_\mathrm{d} \), and returns the filtered signal \( \sigma_\mathrm{ref} \), its first derivative \( \dot{\sigma}_\mathrm{ref} \), and its second derivative \( \ddot{\sigma}_\mathrm{ref} \), corresponding to position-, rate-, and acceleration-level references. The estimated and filtered outputs are sufficiently free of noise~\cite{Grondman2018}. Furthermore, PCH can be readily integrated through the \( \sigma_\mathrm{hdg} \) signal.
The reference model imposes the dynamical behavior on the closed-loop system and thus allows shaping the response and embedding envelope protections. In this work, we impose a limited second-order behavior through \( H_\mathrm{ref}(s) \) on the dynamics similar to~\cite{Milz2022} and shown in~\cref{fig:referencemodel}, though more complex approaches allowing for a more precise shaping exist~\cite{Zhang2013,Bhardwaj2024}:
\begin{subequations}
\begin{equation}\label{eq:pt2_filter}
    H_\mathrm{ref}(s) = \frac{\omega_0^2}{s^2 + 2 \zeta \omega_0 s + \omega_0^2}
\end{equation}
\begin{figure*}[!bht]
    \centering
    \includegraphics[]{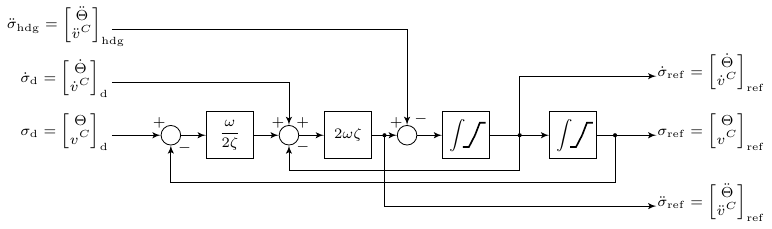}
    \caption{Generic second-order reference model~\cite{Milz2022}.}\label{fig:referencemodel}
\end{figure*}
The subsequent feedback controller minimizes the error of \( \vec{\theta} \) and \( \vec{v}^C \) tracking by and utilized the first- or second-order derivatives from the reference model as a feedforward term. The resulting pseudo control commands are transformed from the control or NED frame into the body frame using kinematic relations~\cite{Milz2026jgcd}.
Additionally, as an absolute heading reference is provided, turn coordination is embedded into the reference model. This allows the heading \( \psi \) to adapt when the roll angle \( \phi \) reference is non-zero during transition or cruise flight. The resulting heading offset \( \psi_\mathrm{coord} \) is described by
\begin{equation}
    \frac{\rm d}{{\rm d}t} \psi_\mathrm{coord} = \frac{g\ \tan\phi_\mathrm{ref}}{\vec{v}^C_\mathrm{x,ref}}
\end{equation}
\end{subequations}
\section{Pilot Control Concept}\label{sec:pilotcontrol}
\begin{figure*}[!htb]
    \centering
    \subfigure[Force-feedback stick mapping\label{fig:inceptor}]
    {
    \includegraphics[]{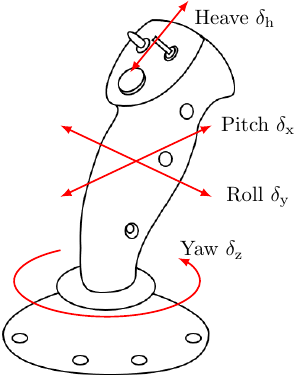}
    }
    \hspace{.1\textwidth}
    \subfigure[Multi-function primary flight display.\label{fig:pfd}]{\includegraphics[width=0.4\textwidth]{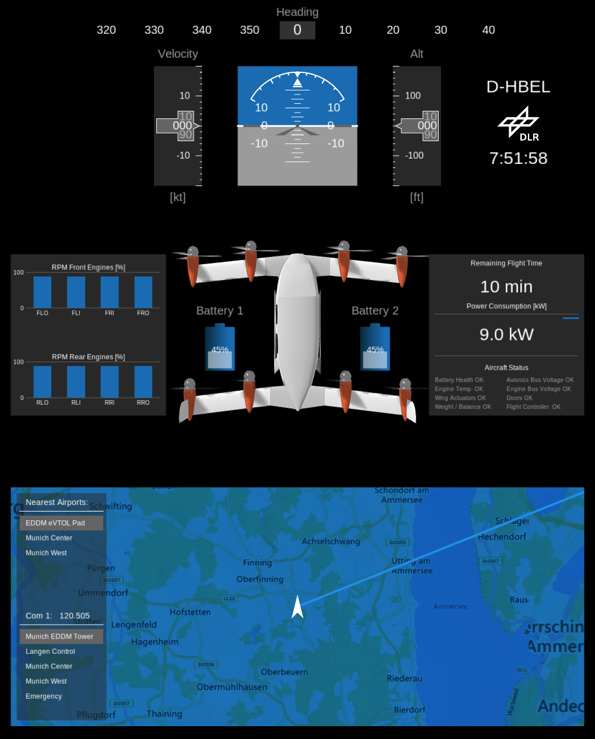}}
    \caption{Cockpit interfaces of the DLR \textit{PAVSIM} cockpit.}\label{fig:hmi_cockpit}
\end{figure*}

Building on the flight-dynamics and control-law framework of \cref{sec:fdm}, this section details how pilot inceptor inputs are converted into phase-consistent commands.
The two-seater cockpit is equipped with a three-axis force-feedback joystick, controlled via EtherCAT, as well as a multi-function touch display (cf.~\cref{fig:pfd}).
The display presents the most relevant flight information, including attitude, altitude, airspeed, and heading. In addition, a virtual top-down camera view was implemented to assist the pilot during landing operations.
The stick is used as the sole input device in the experiments and allows aircraft control in a natural way for untrained pilots~\cite{Born2024}. The side stick and its mapping are shown in~\cref{fig:inceptor}. The environment for the pilots is perceived via Mixed Reality glasses, showing the interior of the cockpit via pass-through cameras and the virtual environment as an overlay replacing the actual windows of the cockpit (see~\cref{fig:dmsc}).
\subsection{Pilot Control Interface}
Based on the mapping of the side stick (\cref{fig:inceptor}) and the controlled variables selection from~\cref{sec:fcinterface}, the pilot control interface can be defined according to the requirements from~\cref{sec:req}. Analogous to~\cref{tab:inceptor_assignments}, \cref{tab:inceptor_assignment} shows the inceptor assignment of the proposed control concept.
The stick signals \( \delta_\mathrm{x} \), \( \delta_\mathrm{y} \), \( \delta_\mathrm{z} \), \( \delta_\mathrm{h} \in \left[ -1; 1 \right] \) are filtered upon entering the controller by applying a dead zone from -0.1 to 0.1, normalized by scaling by \( \frac{1}{0.9} \), and filtered by a first-order low-pass filter of form \( \frac{1}{0.1 s + 1} \), as motivated in~\cite{Fielding2003}. The specific values are within the range suggested in~\cite{DOT_FAA_TC15_13} and tuned to our installed side stick.
\subsection{Command Filters}\label{sec:command_filtering}
\begin{figure}[!htb]
    \centering
    \includegraphics[]{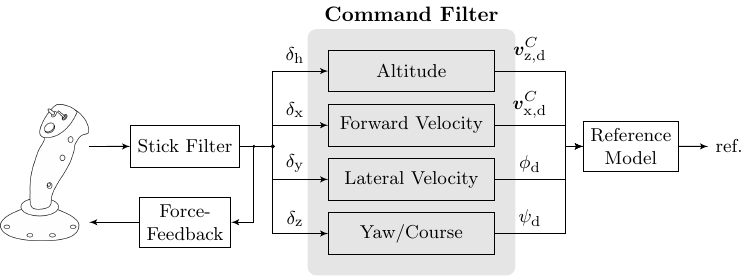}
    \caption{Architecture of the pilot command filter (cf.~\cref{fig:cascade}).}
    \label{fig:comfilter_architecture}
\end{figure}

\begin{figure}[!htb]
    \centering
    \includegraphics[]{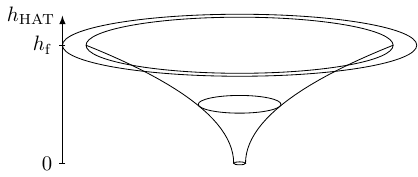}
    \caption{``3D funnel'' over the height above terrain (HAT) \( h_\mathrm{HAT} \) around a vertiport.}
    \label{fig:3dfunnel}
\end{figure}

\begin{figure*}[!htb]
    \centering
    \tikzstyle{block} = [draw, rectangle, minimum height=.7cm, minimum width=.7cm]
    \tikzstyle{input} = [coordinate]
    \tikzstyle{output} = [coordinate]
    \tikzstyle{sum} = [draw, circle, node distance=1cm]
    \resizebox{\textwidth}{!}{\includegraphics[]{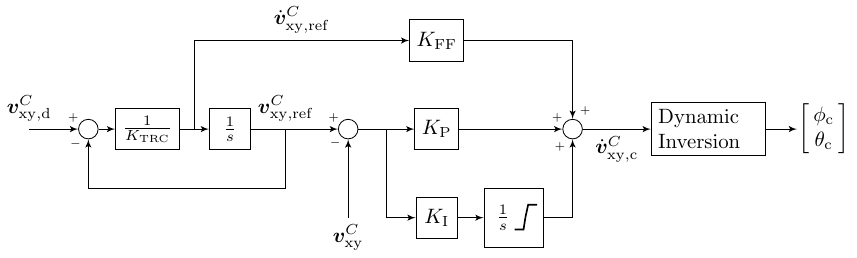}}
    \caption{Block diagram of the translational rate command filter consisting of a first-order reference model, a PI controller, and the inversion block.}\label{fig:translational_rate_control}
\end{figure*}

Command filtering realizes different flight control modes~\cite{Lombaerts2020a,Milz2022}.
The command filtering is separated into command filters for forward and lateral velocity, altitude, and heading. They are based on different mechanisms and concepts including \textit{Hover Damping}, \textit{Height above Terrain (HAT) funnel}, and \textit{Translational Rate Command (TRC)}. The command filters are inspired by~\cite{Lombaerts2020a,Grondman2018,Milz2022} and integrated into the overall control system as shown in~\cref{fig:comfilter_architecture}.
The stick signals are converted by the command filter into controlled-variable commands according to~\cref{tab:inceptor_assignment} to obtain the desired \( \vec{v}^C_\mathrm{d} \) and \( \vec{\theta}_\mathrm{d} \), which are then passed into the reference model. However, these conversions are generally nonlinear, depend on the height above terrain (HAT) \( h_\mathrm{HAT} \) and forward velocity \( \vec{v}^C_\mathrm{x} \), and often involve an intermediate commanded value {\tiny com.}.
Envelope limits are applied throughout on all quantities~\cite{May2026scitech}.
\begin{table*}[!hb]
    \centering
    \caption{Proposed assignment of inceptors and control inputs during the different flight phases written as ``commanded quantity (with hover damping {\tiny dmp.}) \textrightarrow tracked quantity''}\label{tab:inceptor_assignment}
    \begin{tabular}{l|cccc}
        \toprule
        & \phantom{{\tiny (dmp.)} } \( \delta_\mathrm{x} \) & \phantom{{\tiny (dmp.)} } \( \delta_\mathrm{y} \) & \( \delta_\mathrm{z} \) & \( \delta_\mathrm{h} \) \\
        \midrule
        Hover & {\tiny (dmp.)} \( \dot{\vec{v}}^C_\mathrm{x} \to \vec{r}^C_\mathrm{x} \)  & {\tiny (dmp.)} \( \dot{\vec{v}}^C_\mathrm{y} \to \vec{r}^C_\mathrm{y} \) & \( \dot{\psi} \to \psi \) & \( \vec{v}^C_z \to h \) \\
        Transition & \phantom{{\tiny (dmp.)}} \( \dot{\vec{v}}^C_\mathrm{x} \to \vec{v}^C_\mathrm{x} \) & \phantom{{\tiny (dmp.)}} \( \dot{\vec{v}}^C_\mathrm{y} \to \vec{v}^C_\mathrm{y} \) & \( - \to \psi \) & \( \vec{v}^C_z \to h \) \\
        Forward flight & \phantom{{\tiny (dmp.)}} \( \dot{\vec{v}}^C_\mathrm{x} \to \vec{v}^C_\mathrm{x} \) & \phantom{{\tiny (dmp.)}} \( \dot{\vec{v}}^C_\mathrm{y} \to \dot{\psi} \) & \( - \to \chi \) & \( \vec{v}^C_z \to h \) \\
        \bottomrule
    \end{tabular}
\end{table*}
The hybrid NDI controller imposes the reference model on the plant so that the closed-loop properties are inherently tied to the chosen reference dynamics. Consequently, the command filter parameters can be explicitly selected and tuned to target specific natural frequencies and damping ratios derived from standard rotorcraft and fixed-wing handling qualities specifications, mainly the ADS-33~\cite{ads33}. The tuning is done analogously to~\cite{Milz2022}. The final closed-loop properties are assessed through piloted simulations (cf.~\cref{sec:results}).
Hereinafter, the basic concepts applied in the command filters are first described, followed by the actual command filter components, as shown in~\cref{fig:comfilter_architecture}.
\paragraph{Hover Damping}
A crucial point in eVTOL control is the change between flight regimes, particularly between hover and cruise. Each regime has distinct requirements due to their different intuitive means of control. 
In hover, the aircraft behaves like a helicopter and requires precise position control. A zero stick input is expected to maintain stationary hover flight. This involves commanding forward and lateral velocity in the control frame, potentially through TRC, along with yaw rate, and vertical velocity commands.
In cruise, the aircraft emulates fixed-wing characteristics. A zero stick input should maintain a trimmed flight state, while maneuvering is achieved through (rate-commanded) flight-path commands and coordinated turns.
However, these control concepts involve a shift in the order of commanded quantities: velocities are commanded during hover, whereas accelerations are commanded during cruise. To avoid explicit mode switching, we introduce \textit{hover damping}. The method applies acceleration commands permanently but introduces a damping term in hover. This creates an equilibrium for each stick deflection and preserves convergence to stationary hover when the stick is released.
\begin{figure}[!htb]
    \centering
    \includegraphics[]{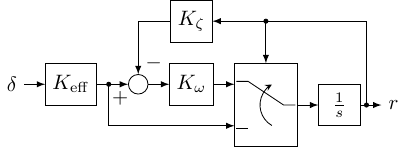}
    \caption{Block diagram of hover damping}
    \label{fig:hover_damping}
\end{figure}
\cref{fig:hover_damping} shows the block diagram of the scheme. A pilot input \( \delta \) is first scaled by a potentially nonlinear effectiveness \(K_\mathrm{eff}\). The reference output is \( r \). Below a threshold value \( r_\mathrm{th} \), a damping term is added to support hover stabilization as the systems naturally stabilizes to the steady hover flight. Above \(r_\mathrm{th}\), the logic transitions to an undamped acceleration command. The filter is initialized with the current state and is defined as:
\begin{subequations}
\begin{equation}\label{eq:hover_damping}
    \frac{\rm d}{{\rm d}t} {r}(t) = \begin{cases}
        K_\mathrm{eff}\, \delta(t) \, , & r(t) > r_\mathrm{th} \\
        K_\omega\,  \left( K_\mathrm{eff}\, \delta(t) - K_\zeta\, r(t) \right)\, , & r(t) \leq r_\mathrm{th}
    \end{cases}
\end{equation}
\begin{figure*}[!htb]
    \centering
    \includegraphics[]{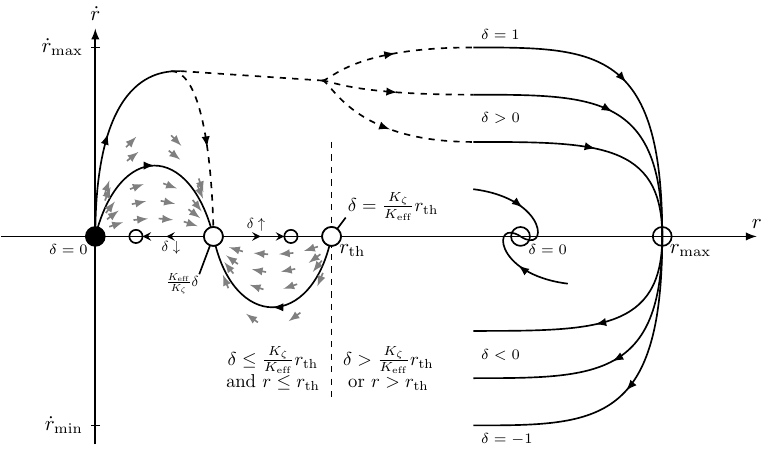}
    \caption{Phase portrait of hover damping with stick input \( \delta \) and command output \( r \).}
    \label{fig:hover_damping_portrait}
\end{figure*}
Using the reference model from~\cref{sec:refmodel}, the phase portrait for the system for different control inputs \( \delta \) is shown in \cref{fig:hover_damping_portrait}.
With \( K_\mathrm{eff} \) and \( K_\zeta \), the stick zone for precise hover flight can be shaped, as equilibria for the system \eqref{eq:hover_damping} are
\begin{equation}
    \begin{cases}
        \delta = 0, & r > r_\mathrm{th} \\
        r = \frac{K_\mathrm{eff}}{K_\zeta}\, \delta, & r \leq r_\mathrm{th}
    \end{cases}
\end{equation}
\end{subequations}
Defining the threshold stick deflection \( \delta_\mathrm{th} = \frac{K_\zeta}{K_\mathrm{eff}} r_\mathrm{th} \) and using the rate limits of the reference for shaping the stick effectiveness allows determining the damping coefficient \( K_\zeta \). \( K_\omega \) can then be used to change the commanded rates below the threshold.
\paragraph{Height above Terrain (HAT) Funnel}
Another crucial filter function is the limitation of the flight envelope, especially the forward and lateral velocity, close to the ground in order to increase safety. This also aligns with flight management requirements, where protected obstacle-free volumes (``3D funnels,'' cf.~\cref{fig:3dfunnel}) are proposed around vertiports~\cite{EASAVertiportPTS2022}.
The height above terrain \( h_\mathrm{HAT} \in \mathbb{R}^+_0 \) is used for the scheduling.
Approaching the ground, the commands and limits are scaled down by a factor \( \eta(h_\mathrm{HAT}) \in \left[ 0; 1 \right] \) to reduce the available envelope:
\begin{equation}
    \eta (h_\mathrm{HAT}) = \left(  \left[ \frac{ h_\mathrm{HAT} }{ h_\mathrm{f} }  \right]^{1}_{0} \right)^2
\end{equation}
with the \textit{funnel height} \( h_\mathrm{f} \) (cf.~\cref{fig:3dfunnel}).
\paragraph{Translational Rate Command (TRC)}
The TRC command filter realizes translational movement in hover flight by changing the vehicle's attitude and thus its thrust vector~\cite{Lombaerts2020a,Milz2022,Patel1998}.
TRC can be viewed as an inversion-based controller that tracks ground-based velocities in horizontal directions, i.e., \( \vec{v}^C_\mathrm{x} \) and \( \vec{v}^C_\mathrm{y} \). 
This allows for precise maneuvering with respect to the ground, especially advantageous when taking off and landing.
The implementation of the TRC is shown in detail in~\cite{Lombaerts2020a,Milz2022}. 
However, this work only utilizes TRC for lateral control (\( \vec{v}^C_\mathrm{y} \)), as the tilt wings can control the forward velocity. However, the complete TRC is introduced within this section as other eVTOL may not posses this capability and thus require TRC for forward flight.
The block diagram is sketched in \cref{fig:translational_rate_control}.
The TRC filter can be expressed as
\begin{subequations}
\begin{equation}\label{eq:trc_filter}
    \dot{\vec{v}}_\mathrm{xy,c}^{C} = \left( K_\mathrm{P} + \frac{K_\mathrm{I}}{s} \right) \left( \frac{1}{K_\mathrm{TRC} s+1} {\vec{v}_\mathrm{xy,d}}^{C} - {\vec{v}_\mathrm{xy}}^{C} \right) + K_\mathrm{FF} \frac{s}{K_\mathrm{TRC} s+1} {\vec{v}_\mathrm{xy,d}}^{C}
\end{equation}
The dynamic inversion is performed according to the following law~\cite{Milz2022}
    \begin{align}
        \phi_\mathrm{c}   & = \arcsin \left( \frac{ \dot{\vec{v}}^C_\mathrm{y,c} }{ g - \dot{\vec{v}}^C_\mathrm{z,c} } \right) \\
        \theta_\mathrm{c} & = \arctan \left( \frac{ \dot{\vec{v}}^C_\mathrm{x,c} \cdot \cos \phi \tan \delta_\mathrm{tilt} + \dot{\vec{v}}^C_\mathrm{z,c} - g}{ \dot{\vec{v}}^C_\mathrm{x,c} + \left( \dot{\vec{v}}^C_\mathrm{z,c} - g \right)  \cos \phi \tan \delta_\mathrm{tilt}} \right)
    \end{align}
\end{subequations}
The linear controller consists of a feed-forward path with a PI feedback controller.
A proportional control cascade is added around the velocity controller for position control capabilities of the TRC.
\subsubsection{Altitude Command Filter}
The altitude command filter controls the altitude by commanding a vertical velocity \( \vec{v}^C_\mathrm{z} \). At sufficient HAT, the heave button commands discrete climb/descent rates, which change throughout the envelope. If the vehicle approaches the ground, the descent rate is decreased to provide a soft touchdown. The command filter is described by the following equations
\begin{subequations}
    \begin{align}
        \dot{h}_\mathrm{com} &= \delta_\mathrm{h} \cdot \begin{cases}
            \dot{h}_\mathrm{max} \ , & \delta_\mathrm{h} \geq 0  \\
            \dot{h}_\mathrm{min} \, \eta(h_\mathrm{HAT}) \, \eta(\dot{\vec{v}}^C_\mathrm{x,d}-\dot{\vec{v}}^C_\mathrm{x,min}) \ , & \delta_\mathrm{h} < 0
        \end{cases} \\
        \frac{\rm d}{{\rm d}t} h_\mathrm{com}(t) &= \dot{h}_\mathrm{com} \\
        \vec{v}^C_\mathrm{z,d} &= - K_h \left( h_\mathrm{com} - h \right)  \\
        \dot{\vec{v}}^C_\mathrm{z,d} &= - K_{\dot{h}} \left( \dot{h}_\mathrm{com} + \vec{v}^C_\mathrm{z} \right)
    \end{align}
\end{subequations}
with the parameters for the maximum/minimum climb/sink rate \( \dot{h}_\mathrm{max} \) and \( \dot{h}_\mathrm{min} \) and the time constant for the altitude rate command \( {K_{\dot{h}} = \num{0.2}} \) as well as the altitude feedback controller \( {K_h = \num{0.5}} \). 
The rate commands are transformed with a piecewise linear function since the absolute maximum rate is lower than the minimum rate. The term {\( { \eta(\dot{\vec{v}}^C_\mathrm{x,d}-\dot{\vec{v}}^C_\mathrm{x,min}) } \)} works similar to the 3D funnel, but reduces the maximum sink rate when decelerating as energy constraints hold~\cite{May2025jgcd}.
\subsubsection{Forward Velocity Command Filter} 
The velocity command filter has a similar basic structure to the altitude command filter. However, the maximum allowable horizontal velocity decreases with lower HAT due to the 3D funnel. For the hover damping, the stick threshold is selected as \( {\delta_\mathrm{th} = \num{0.7}} \), i.e., the value for which continuous deflections leads to the transition from hover mode, to allow for precise hovering.
The command filter is defined as
\begin{subequations}
\begin{align}
    \dot{\vec{v}}^{C}_\mathrm{x,com} &= \delta_\mathrm{x} \, \eta(h_\mathrm{HAT}) \, \begin{cases}
        \dot{v}^C_\mathrm{x,min} \ , & \delta_\mathrm{x} < 0  \\
        \dot{v}^C_\mathrm{x,max} \ , &  0 \leq \delta_\mathrm{x}
    \end{cases} \\
    \frac{\rm d}{{\rm d}t} \vec{v}^C_\mathrm{x,d}(t) &= \begin{cases}
            \dot{\vec{v}}^{C}_\mathrm{x,com}, & \vec{v}^C_\mathrm{x,d} > V_\mathrm{h/t} \\
            K_\mathrm{\omega,v_x} \left( \dot{\vec{v}}^{C}_\mathrm{x,com} - K_\mathrm{\zeta,v_x}\, v^C_\mathrm{x,d} \right), & \mathrm{otherwise} 
    \end{cases}
\end{align}
\end{subequations}
with the minimum and maximum forward accelerations \( \dot{v}^C_\mathrm{x,min} \) and \( \dot{v}^C_\mathrm{x,max} \), and the hover damping gains \( K_\mathrm{\omega,v_x} = \num{3}  \) and \( K_\mathrm{\zeta,v_x} = \frac{\dot{v}^C_\mathrm{x,max}\, \delta_\mathrm{th}}{V_\mathrm{h/t}} \).
For backward flight, velocity command is always on. An outer position-hold loop can be added for improved position accuracy.
\subsubsection{Lateral Velocity Command Filter}
In contrast to the forward velocity, the lateral velocity command interpretation changes throughout the envelope. While in hover, a sidewards stick deflection (\( \delta_\mathrm{y} \)) results in a sidewards velocity, induced by a roll angle for the tilt-wing configuration. However, during cruise, no direct lateral velocity can be controlled, but a roll angle inducing a (coordinated) turn maneuver. 
Both effects can be uniformly combined in the lateral acceleration or load factor in the control frame, i.e., \( \dot{\vec{v}}^C_\mathrm{y} \) or \( n^C_\mathrm{y} \), although the coordinated turn does not change the velocity in the control frame \( \vec{v}^C \), as the frame rotates with the heading.
\begin{figure}[!htb]
    \centering
    \includegraphics[]{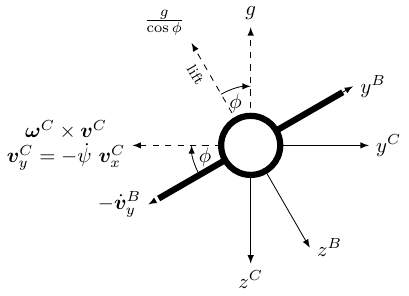}
    \caption{Sketch of the aircraft during a coordinated turn.}
    \label{fig:coordinated_turn}
\end{figure}
\cref{fig:coordinated_turn} sketches the frames and forces of the vehicle in a coordinated turn. In the control frame, a fictitious force resulting from kinematic constraints acts on the vehicle. The sidewards load factor resulting from tilting the upward force (lift and/or thrust) \( \vec{n}^C_\mathrm{y,sd} \) can be approximated as
\begin{subequations}
\begin{align}
    \vec{n}^C_\mathrm{y,sd} = \sin\phi\, \vec{n}^B_\mathrm{z} = \tan\phi\, \vec{n}^C_\mathrm{z} 
\end{align}
because \( \vec{n}^B_y = 0 \) during a coordinated turn, we can assume \( \vec{n}^B_z \approx \frac{\vec{n}^C_z}{\cos\phi} \) for small pitch angles.
While during hover the lateral motion is counteracted by drag force, leading to an equilibrium, during a coordinated turn it is opposed by the centripetal force or load factor \( \vec{n}^V_\mathrm{y,cp} \), i.e.,
\begin{equation}
    \vec{n}^C_\mathrm{y,cp} = \frac{\dot{\psi}\ \vec{v}^C_\mathrm{x}}{g}
\end{equation}
with the yaw rate \( \dot{\psi} \), leading to
\begin{equation}
    \vec{n}^C_\mathrm{y} = \vec{n}^C_\mathrm{y,sd} + \vec{n}^C_\mathrm{y,cp} = \tan\phi\, \vec{n}^C_\mathrm{z} + \frac{\dot{\psi}\ \vec{v}^C_\mathrm{x}}{g}
\end{equation}
This allows to realize \( \vec{n}^C_y \) through a constant turn rate \( \dot{\psi} \) if sufficient forward velocity is present, or by solely tilting the upward force vector in hover.
Additionally, HAT scaling is applied to the sideways acceleration command, but not to the coordinated turn command, since the latter is assumed to occur at a sufficiently high altitude.
Solving for the commanded roll angle \( \phi_c \) yields
\begin{equation}
    \phi_c = \arctan \frac{\eta(h_\mathrm{HAT})\, \vec{n}^C_{y,c} - \frac{\dot{\psi}_c \ \vec{v}^C_x}{g}}{n^C_z} = \arctan \frac{ \eta(h_\mathrm{HAT})\, \dot{\vec{v}}^C_{y,c} - \dot{\psi}_c \vec{v}^C_x }{ g + \dot{\vec{v}}^C_z }
\end{equation}
where the turn rate command \( \dot{\psi}_c \) equals
\begin{equation}\label{eq:phi_dot_turn}
    \dot{\psi}_c = \begin{cases}
        0, & \vec{v}^C_x \leq V_\mathrm{h/t} \\
        \frac{\vec{n}^C_{y,c} \ g}{\vec{v}^C_x} = \frac{ \dot{\vec{v}}^C_{y,c} }{\vec{v}^C_x}, & \mathrm{otherwise}
    \end{cases}
\end{equation}
leading to
\begin{equation}
    \phi_d = \phi_c + \arctan\frac{\dot{\psi}_c\ \vec{v}^C_x}{g}
\end{equation} 
\end{subequations}
\subsubsection{Yaw/Course Command Filter} 
The yaw control filter implements a yaw rate command and direction or course hold logic.
While~\cite{Lombaerts2020a} uses the yaw rate in body axes \( r \), the derivative of the heading in earth axes \( \dot{\psi} \) is used in this filter~\cite{Milz2022}. While during hover the heading \( \psi \) is tracked, during cruise flight, the flight path course angle \( \chi^K \) will be tracked using the wind‑correction angle \( \beta^K \)
\begin{subequations}
\begin{align}
    \dot{\psi}_\mathrm{com} &= \dot{\psi}_\mathrm{max} \ \left( \frac{ \vec{v}^C_x  }{ V_\mathrm{h/t} } - 1 \right)^2 \ \delta_\mathrm{z} \\
    \frac{\rm d}{{\rm d} t} \psi_\mathrm{d}^\prime &=  \dot{\psi}_\mathrm{com}(\tau) \\
    \psi_\mathrm{d} &= \psi_\mathrm{d}^\prime + \left[ \frac{\vec{v}^C_\mathrm{x} - V_\mathrm{h/t}}{V_\mathrm{t/c} - V_\mathrm{h/t}} \right]^1_0 \beta^K
\end{align}
\end{subequations}
The effectiveness of the yaw command decreases with velocity and vanishes when entering the transition regime. During cruise, a change in course is realized through lateral stick inputs (roll), as described in~\eqref{eq:phi_dot_turn}, which must be considered in the reference heading.
\subsection{Force-Feedback}
The force-feedback joystick can be configured with two parameters per axis (longitudinal \( x \), lateral \( y \), yaw \( z \)), i.e., the zero point \( \delta_0 \in \left[ -1; 1 \right] \) and resistance \( R \in \left[ 0; 1 \right] \), where already some base resistance is applied.
The zero point is calculated to guide the pilot toward (energy-)optimal flight points and thus ``push'' through the transition and into level flight.
The lateral axis thus uses the roll angle \( \phi \) to set the zero point of the stick, gently guiding the pilot into level flight for small roll angles. 
For the lateral axis, the neutral point and the resistance depend on a limit angle \( \phi_\mathrm{lim} < \phi_{\max} \) and are defined as:
\begin{subequations}
\begin{align}
    \delta_{0,\rm y} &= \begin{cases}
         -\mathrm{sign}\phi \frac{ \phi_\mathrm{lim} - \phi }{ \phi_{\max} - \phi_\mathrm{lim} }, & \vec{v}^C_\mathrm{x} > V_\mathrm{h/t} \land \left| \phi \right| > \phi_\mathrm{lim} \\
        0, & \mathrm{otherwise}
    \end{cases} \\
    R_\mathrm{y} &= \begin{cases}
        \frac{\left| \phi \right|}{\phi_\mathrm{lim}}\, ,& \left| \phi \right| \leq \phi_\mathrm{lim} \\
        1 \, ,& \mathrm{otherwise}
    \end{cases}
\end{align}
\begin{figure}[!htb]
    \centering
    \includegraphics[]{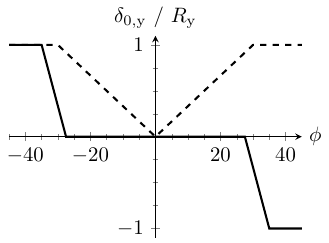}
    \hfill
    \includegraphics[]{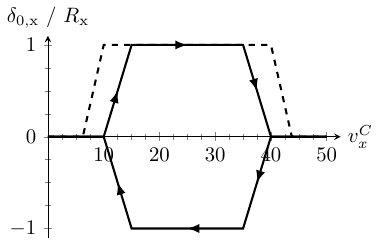}
    \caption{Active side stick neural point \( \delta_0 \) (\tikzline{}) and resistance \( R \) (\tikzline{dashed}) for longitudinal \( y \) and lateral \( x \) axis}
    \label{fig:force_feedback_plot}
\end{figure}
For the longitudinal axis, the horizontal velocity \( \vec{v}^C_\mathrm{x} \) is used to guide the pilot through the transition phase. The guidance blends in in a region of \( \Delta V_\mathrm{ffb} \) at the borders of the transition phase. The neutral point and the resistance are:
\begin{align}
    \delta_{0,\rm x} &= \mathrm{sign}\, \dot{\vec{v}}^C_\mathrm{x,c}\ \begin{cases} 
        \frac{\vec{v}^C_x - V_\mathrm{h/t}}{\Delta V_\mathrm{ffb}}, & V_\mathrm{h/t} \leq \vec{v}^C_x < V_\mathrm{h/t} + \Delta V_\mathrm{ffb} \\
        1 , &  V_\mathrm{h/t} + \Delta V_\mathrm{ffb} \leq v^C_x \leq V_\mathrm{t/c} - \Delta V_\mathrm{ffb} \\
        \frac{V_\mathrm{t/c} - \vec{v}^C_x}{\Delta V_\mathrm{ffb}}, & V_\mathrm{t/c} -  \Delta V_\mathrm{ffb} < \vec{v}^C_x \leq V_\mathrm{t/c} \\
        0, & \mathrm{otherwise}
    \end{cases} \\
    R_\mathrm{x} &= \begin{cases}
        1 + \frac{\vec{v}^C_\mathrm{x} - V_\mathrm{h/t}}{ \Delta V_\mathrm{ffb} }\, , & V_\mathrm{h/t} -  \Delta V_\mathrm{ffb} \leq \vec{v}^C_\mathrm{x} < V_\mathrm{h/t} \\
        1   \, , & V_\mathrm{h/t} \leq \vec{v}^C_\mathrm{x} < V_\mathrm{t/c} \\
        \frac{V_\mathrm{t/c} - \vec{v}^C_\mathrm{x} }{ \Delta V_\mathrm{ffb} } \, , & V_\mathrm{t/c} - \Delta V_\mathrm{ffb} \leq \vec{v}^C_\mathrm{x} < V_\mathrm{t/c} \\
        0 \, , & \text{otherwise}
    \end{cases}
\end{align}
\end{subequations}
\section{Validation and Results}\label{sec:results}
After presenting the command-filter design, we evaluate the concept in a staged validation workflow that moves from controller-centric evidence to human-in-the-loop evidence.
The proposed control concept is validated using the closed-loop tandem tilt-wing simulation model~\cite{Milz2026jgcd}. The validation is threefold:
First, simulation results of a mission are used to show the vehicle's response to inceptor inputs.
Then, pilot-in-the-loop simulations are performed on the DLR \textit{PAVSIM} simulator with an active side stick, where both the system's responses as well as subjective pilot ratings are collected.
Finally, an optimal-control-based analysis of the concept is performed.
\subsection{Simulative Results}\label{sec:res:sim}
We evaluate the concept on a simplified mission profile. The vehicle starts from ground hover. It climbs to approximately \SI{200}{\meter}. Once \SI{30}{\meter} altitude is exceeded, forward acceleration is added to reach cruise flight quickly.
In cruise, a \SI{90}{\degree} coordinated turn is executed. The vehicle then descends to \SI{100}{\meter}. From that point, deceleration and descent are applied simultaneously. After re-entering hover, the vehicle lands smoothly.
\begin{figure*}[!htb]
    \centering
    \includegraphics[]{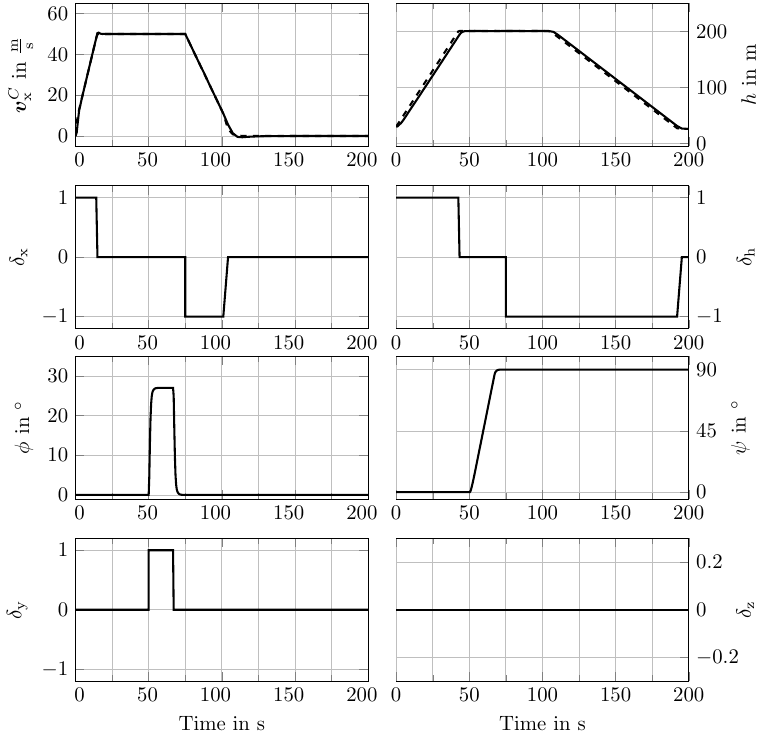}
    \caption{Forward velocity \( \vec{v}^C_\mathrm{x} \), altitude \( h \), roll angle \( \phi \), heading \( \psi \), load factor \( \vec{n}^B \), and inceptors \( \delta \) for the simulative case with commanded (\tikzline{dashed}) and actual (\tikzline{}) signals.}\label{fig:results_sim}
\end{figure*}
The results are shown in~\cref{fig:results_sim}. Overall trends track the command profiles smoothly, including the accelerating and decelerating transitions. The mission also requires only a simple inceptor-input sequence, which supports the intended low-complexity interaction concept.
The altitude trend shown in~\cref{fig:results_sim} reflects the lack of a feedforward component in the altitude command filter as the actual altitude trend follows the reference and settles to the final value without an overshoot.
\subsection{Pilot-in-the-Loop on DLR \textit{PAVSIM} Motion Simulator}\label{sec:res:pavsim}
\begin{figure*}[!htb]
    \centering
    \includegraphics[width=.9\textwidth,trim=3cm 5cm 10cm 0,clip]{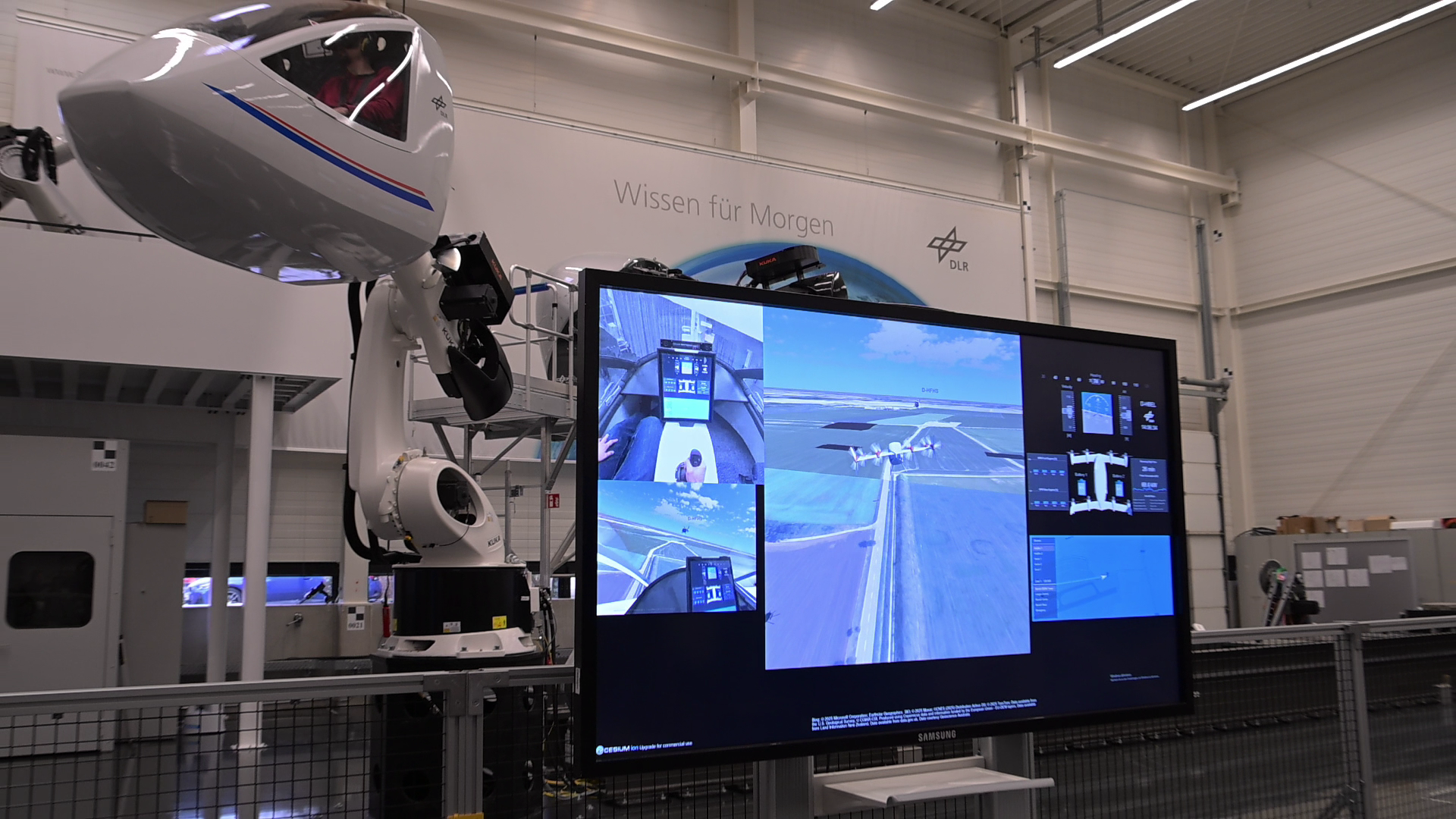}
    \caption{DLR's Dynamic Motion Simulation Center showing the PAVSIM simulator with live view from the cockpit interior, display, and the pilot's augmented reality.}
    \label{fig:dmsc}
\end{figure*}

\begin{figure*}[!htb]
    \includegraphics[]{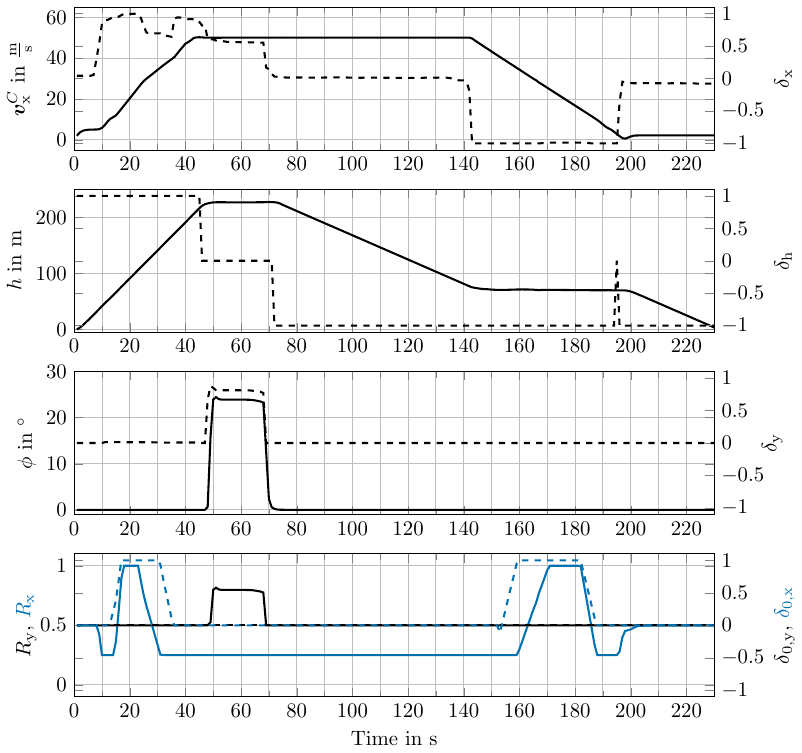}
    \caption{Forward velocity \( \vec{v}^C_\mathrm{x} \), altitude \( h \), roll angle \( \phi \), heading \( \psi \), and inceptor inputs \( \delta \) for the optimal-control cases with signals (\tikzline{}) and inceptor (\tikzline{dashed}) with a base resistance of \num{0.25}.}\label{fig:results_pavsim}
\end{figure*}

The control concept is implemented and validated on a motion simulator platform, the personal air vehicle simulator \textit{PAVSIM} shown in~\cref{fig:dmsc}.
It is an experimental research flight simulator developed by DLR to investigate future aviation concepts and advanced cockpit technologies. The simulator consists of a two-seat cockpit mounted on an industrial robot, enabling large-amplitude motions and highly dynamic motion cueing beyond the capabilities of conventional hexapod simulators.
This configuration allows researchers to study flight dynamics, control strategies, human–machine interaction, pilot workload, and immersive cockpit environments under realistic motion conditions~\cite{Seefried2019}.
The robot motion is generated using a one-step-optimization-based motion cueing algorithm~\cite{Bellmann2011}. 
The cockpit is equipped with two seats next to each other, sharing a multifunctional touch display and a force-feedback side stick (see~\cref{fig:hmi_cockpit}).
For validation, the proposed control concept has been implemented and tested on the motion simulator.
\cref{fig:results_pavsim} show the recorded data of one flight that represents a similar maneuver as in~\cref{sec:res:sim}.
First qualitative pilot feedback indicates that non-expert participants adapted quickly to the mapping, while conventionally trained pilots required additional familiarization with the non-traditional inceptor semantics.
The results in~\cref{fig:results_pavsim} additionally show that also in pilot-in-the-loop tests, only little stick activity is required and applied during the generic mission.
\subsection{Optimal-Control-based Testing}\label{sec:res:ocp}
This subsection complements the simulation and pilot-in-the-loop results with an optimization-based performance reference. Two use cases are considered: (i) quantifying the performance penalty introduced by pilot command filters compared with direct control-input optimization, and (ii) investigating the inceptor movements required on baseline maneuvers in piloted flight.
The optimal-control problem seeks controls $ \vec{u}\!\left( t \right)$,  states $ \vec{x} \!\left( t \right) $, and terminal time $t_f$ for the nonlinear system \(\dot{\vec{x}}\!\left(t\right) = \vec{f} \! \left(t,\vec{x}\!\left(t\right),\vec{u}\!\left(t\right)\right)\) that brings the system from its initial state $\vec{x}_0 = \vec{x}\!\left(t_0\right)$ to a desired terminal state at $\vec{x}_f = \vec{x}\!\left(t_f\right)$ with minimized cost $ J(\vec{x},\vec{u},t_f)$ while complying with path and derivative constraint functions.
The continuous-time problem is transcribed using direct collocation with trapezoidal integration on an equidistant grid of 60 nodes per phase, where all discretized states, inputs, and $t_f$ are optimization variables (cf.~\cref{tab:Vars}). The applied framework is introduced and described in more detail in~\cite{May2025jgcd}.
The resulting trajectories are interpreted as best-case references for mission time or input reduction under the imposed model and constraints; they are not interpreted as direct proxies for pilot performance.
In order to compare both, the potential performance penalty introduced by the command filter as well as the input-reduction through the force feedback, we propose to investigate five cases within one scenario:
\begin{enumerate}\setcounter{enumi}{-1}
    \item Open-loop aircraft as a baseline and lower bound for the comparison. However, because there are few limits imposed compared to the envelope-protected controller, significantly faster maneuvers can be achieved.
    \item Closed-loop aircraft, where the input is directly fed into the reference model. This case gives a lower bound to what is possible while obeying the performance and envelope limitations given by the reference model.
    \item Full simulation with command filters and inceptor inputs, with the goal of fulfilling the mission as fast as possible and
    \begin{enumerate}
        \item accepting every possible input.
        \item penalizing control inputs.
        \item penalizing control inputs but using force feedback to deflect the stick's neutral point autonomously.
    \end{enumerate}
\end{enumerate}
The cost functions \( J  \) for each case are:
\begin{subequations}
\begin{align}
    J_\mathrm{0/1/2a}(t_f, \vec{x}, \vec{u}) &= t_f \\
    J_\mathrm{2b/2c}(t_f, \vec{x}, \vec{u}) &= t_f + \int_0^{t_f} \sum_{\mathclap{\qquad i \in \left\{ x,y,z,h \right\}}} \left( \delta_i - \delta_{0,i} \right)^2 \ {\rm d}t
\end{align}
\end{subequations}
The scenario is defined analogous to the simplified mission profile from~\cref{sec:res:sim} but neglecting the pure hover climb and sink phases. It is implemented as a 3-phase optimal control problem: (i) accelerating climbing transition, (ii) coordinated turn in cruise flight, and (iii) decelerating sinking transition. The inputs and states of the problem are shown in \cref{tab:Vars}. The four boundary conditions of the three phases on the flight dynamics state vector are
\begin{align*}
    x_0 &= \left\{ \vec{r}^E_\mathrm{z} = \SI{-30}{\meter},\ \vec{v}^C = \vec{\theta} = \vec{\omega}^B = 0 \right\} \\
    x_1 &= \left\{ \vec{r}^E_\mathrm{z} = \SI{-200}{\meter} ,\ \vec{v}^C = \left[ 50\ 0 \ 0 \right]^T \si{\meter\per\second}, \phi = \psi = 0 \right\} \\
    x_2 &= \left\{ \vec{r}^E_\mathrm{z} = \SI{-200}{\meter},\ \vec{v}^C = \left[ 50\ 0 \ 0 \right]^T \si{\meter\per\second}, \phi = 0, \psi = \SI{90}{\degree} \right\} \\
    x_3 &= \left\{ \vec{r}^E_\mathrm{z} = \SI{-30}{\meter},\ \vec{v}^C = \left[ 0\ 0 \ 0 \right]^T \si{\meter\per\second}, \phi = 0, \psi = \SI{90}{\degree} \right\}
\end{align*}
\begin{figure*}[!htb]
        \definecolor{color1}{HTML}{0072B2}
        \definecolor{color2}{HTML}{E69F00}
        \definecolor{color3}{HTML}{56B4E9}
        \includegraphics[]{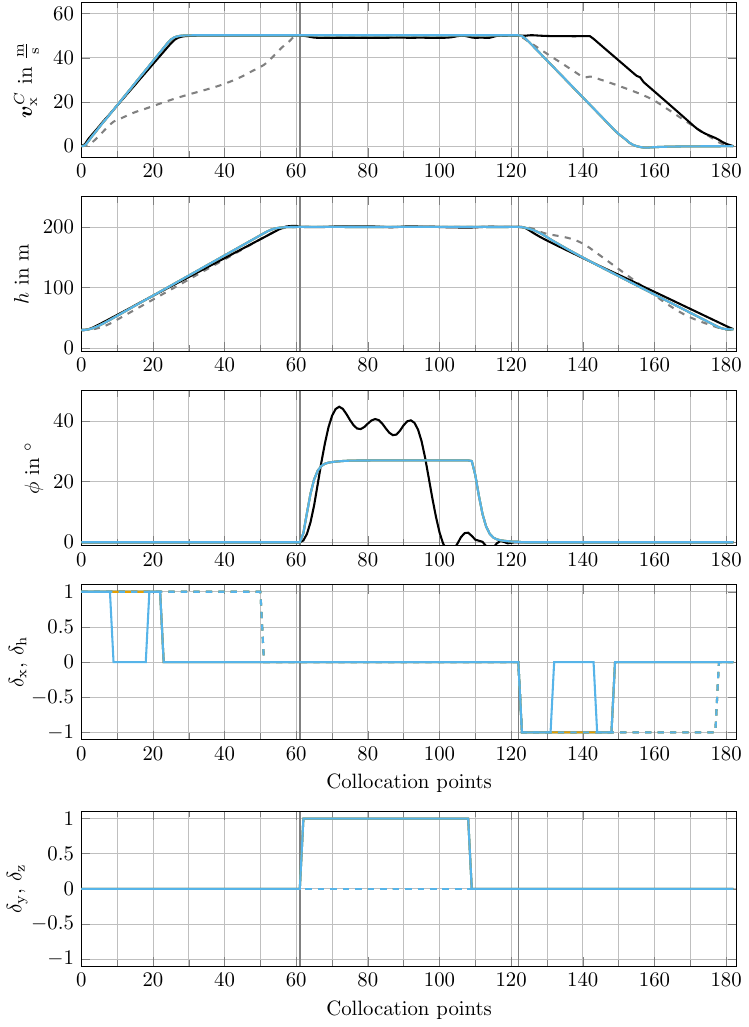}
    \caption{Forward velocity \( \vec{v}^C_\mathrm{x} \), altitude \( h \), and roll angle \( \phi \) with commanded (\tikzline{dashed}) and actual (\tikzline{}) signals, and inceptor inputs \( \delta_\mathrm{x} \), \( \delta_\mathrm{y} \) (\tikzline{}), and \( \delta_\mathrm{h} \), \( \delta_\mathrm{z} \) (\tikzline{dashed}) for the optimal-control cases 1 (\tikzline{}), 2(a) (\tikzline{color1}), 2(b) (\tikzline{color2}), 2(c) (\tikzline{color3}).}\label{fig:results_ocp}
\end{figure*}
The results of the optimization are shown in~\cref{fig:results_ocp}. The time line of the plots is normalized, and the final times of each case are: \( t_\mathrm{f,1} = \SI{113.3}{\second} \), \( t_\mathrm{f,2a} = \SI{127}{\second} \),
\( t_\mathrm{f,2b} = \SI{127}{\second} \) (\( J_\mathrm{2b} = \num{281.8}\)),
and \( t_\mathrm{f,2c} = \SI{127}{\second} \) (\( J_\mathrm{2c} = \num{261.7}\)).
For reference, the open-loop case is added, which has \( t_\mathrm{f,0} = \SI{58.1}{\second} \).
\section{Discussion}
This section discusses the three validation steps with respect to the requirements in \cref{sec:req}.
To preserve the current flight state after stick release while still enabling precise hover flight, the hover-damping concept is introduced. This in combination with a decoupling control law, i.e., dynamic inversion, allows all phases to be handled without explicit mode switching. 
Transparency is further supported by the unified mapping and reinforced by force-feedback side stick.
The simulation results shown in \cref{sec:res:sim} indicate that the aircraft can be operated across and transition between all flight phases stably.
Furthermore, they show that the command filters deliver smooth, low-jerk trajectories across hover, transition, and cruise flight.
The hover-damping filter proves to maintain the commanded quantities as expected.
However, \cref{fig:results_sim} shows a deceleration change when entering the hover regime. Although the effect is amplified in the shown case because the inceptor is released simultaneously and force-feedback damping is absent, this behavior remains a drawback of the current hover-damping formulation.
Pilot-in-the-loop tests on the DLR PAVSIM (Sec.~\ref{sec:res:pavsim}) investigate and show the functionality of the proposed control concept on a motion simulator platform. This represents a main step towards further studies on the control concept.
The results suggest that the force-feedback side stick can guide pilots through transitions. 
At the current stage, these human-factors findings should be interpreted as preliminary because the available evidence is primarily qualitative.
Interestingly, initial feedback on the single-inceptor mapping indicates that non-expert users may familiarize themselves faster with the concept than conventionally trained pilots. A similar study and observations are shown in~\cite{Jones2014}.
However, stronger conclusions regarding pilot-in-the-loop metrics require a larger and statistically powered study.
A systematic comparison between the optimal trajectories obtained via direct collocation and the trajectories generated by the command filters allows us to quantify the residual performance penalty of the proposed control concept. Compared with the closed-loop aircraft case, where direct commands are issued, the control inceptor slows the overall mission (by around \SI{12}{\percent}). This is mainly caused by a more strict decoupling of the axis which does not allow more coupled maneuvers to reduce the mission time. However, this comes at the advantage of more intuitive and more comfortable maneuvers. This can also be seen when comparing with the results from~\cref{sec:res:sim,sec:res:pavsim}.
Furthermore, it is interesting to note that the optimal responses in the roll angle are similar and dictated by the controller, more precisely, the reference model (limits). Only the open-loop case is free to exceed the limit, leading to a significantly accelerated turn, with a reduction from \SI{20}{\second} to \SI{12.3}{\second}.
This, however, is achieved only with large turn rates.
Regarding longitudinal motion (\( \vec{v}^C_\mathrm{x} \) and \( h \)), the command filter exhibits little coordination between the two quantities, primarily due to the limitations of the single side-stick and the discrete vertical motion control. Consequently, both the open-loop and closed-loop cases coordinate maneuvers differently to expedite both the accelerating (open-loop: \SI{19.2}{\second}, closed-loop: \SI{37.5}{\second}, command filtered: \SI{43}{\second}) and decelerating (open-loop: \SI{26.1}{\second}, closed-loop: \SI{57}{\second}, command filtered: \SI{67}{\second}) transitions.
Finally, an analysis of the inceptor inputs reveals they are discrete, as the fastest mission times are achieved by pushing the inceptors to their limits. Furthermore, the reduction in \( \delta_\mathrm{x} \) input during both transition phases is notable.
These findings also strengthen and validate the optimal control-based trajectory-based analysis of the command filters. Although this validation using single trajectories is not fully representative, it already provides a first indication. Another approach would be to employ attainable moment sets~\cite{Lu2023} of the vehicle and analyze its shrinking when adding the command filter.
In conclusion, the unified pilot-control scheme presented herein meets the specified requirements, operates reliably across all flight regimes, and provides an intuitive interface that benefits both trained and amateur pilots. 
\section{Conclusion}
This study presented the design, implementation, and initial validation of a unified pilot control concept tailored for transformational electric vertical take-off and landing (eVTOL) aircraft, demonstrated on a tandem tilt-wing configuration.
The proposed architecture combines SVO-oriented command filtering with a hybrid nonlinear dynamic inversion control law to support continuous operation across all flight phases without mode switching. The underlying control architecture effectively decouples the aircraft's dynamics and serves as the basis for unified control.
Simulation results show stable execution of representative multi-phase maneuvers and indicate smooth command behavior under the tested conditions, allowing the pilot to follow appropriate mission profiles with simple command inputs while providing a smooth ride.
The concept, together with the active force-feedback side stick, enhances intuitive control without requiring mode switching or complex coordination and has the potential to reduce pilot workload.
Pilot-in-the-loop experiments on the DLR PAVSIM robotic motion simulator provide encouraging preliminary evidence that the active force-feedback side stick supports multi-phase flight capabilities, ensures transparent control, and improves concept learnability.
First results indicate that the control interface is particularly effective for untrained pilots, supporting safe operations in low-altitude economy contexts by aligning with regulatory requirements for controllability and flying qualities.
Optimal-control-based analysis of the control concepts suggests that the active side stick reduces required inceptor inputs throughout the mission. While it further reveals a penalty in mission time compared to an optimally driven closed-loop system, it enhances overall maneuver smoothness.
Ultimately, this unified control strategy proves to be a viable and effective method for managing the complex dynamics of eVTOL vehicles.
\clearpage
\backmatter
\begin{appendices}
\section{Optimal Control Problem Variables}\label{secA1}
\begin{table}[!htb]
\centering
\begin{tabular}{ c c c c c c } 
    \toprule
    Variable & Unit & Lower Bound & Upper Bound & Lower RC $\left(\frac{\mathrm{Unit}}{\si{\second}}\right)$ & Upper RC $\left(\frac{\mathrm{Unit}}{\si{\second}}\right)$ \\ 
    \midrule
    \multicolumn{6}{ c }{State Variables} \\
    \midrule
    $\vec{r}^E_\mathrm{x}$ & $\si{\meter}$ & \num{-1e4} & \num{1e4} & \num{-60} & \num{60} \\ 
    $\vec{r}^E_\mathrm{y}$ & $\si{\meter}$ & \num{-1e4} & \num{1e4} & \num{-60} & \num{60} \\ 
    $\vec{r}^E_\mathrm{z}$ & $\si{\meter}$ & \num{-220} & \num{0}   & \num{-10}  & \num{10}  \\
    $\vec{v}^B_\mathrm{x}$ & $\si{\meter\per\second}$ & \num{-5} & \num{55} & \num{-2.5} & \num{5} \\
    $\vec{v}^B_\mathrm{y}$ & $\si{\meter\per\second}$ & \num{-5} & \num{5} & \num{-2.5} & \num{2.5} \\
    $\vec{v}^B_\mathrm{z}$ & $\si{\meter\per\second}$ & \num{-10} & \num{10} & \num{-5} & \num{5} \\
    $\phi$   & $\si{\degree}$ & \num{-45} & \num{45} & \num{-30} & \num{30} \\
    $\theta$ & $\si{\degree}$ & \num{-15} & \num{15} & \num{-5} & \num{5} \\
    $\psi$   & $\si{\degree}$ & -M & M & \num{-15} & \num{15} \\
    $\vec{\omega}^B_\mathrm{x}$ & $\si{\degree\per\second}$ & \num{-30} & \num{30} & \num{-10} & \num{10} \\ 
    $\vec{\omega}^B_\mathrm{y}$ & $\si{\degree\per\second}$ & \num{-15} & \num{15} & \num{-10} & \num{10} \\ 
    $\vec{\omega}^B_\mathrm{z}$ & $\si{\degree\per\second}$ & \num{-15} & \num{15} & \num{-10} & \num{10} \\
    $\vec{\delta}_{w}$ & $\si{\degree}$ & \num{0} & \num{90} & \num{-30} & \num{30} \\
    $T_\Sigma$ & $\si{\newton}$ & \num{0} & \num{10e3} & \num{-1e3} & \num{1e3} \\
    $T_{\Delta x}$ & $\si{\newton}$ & \num{-5e2} & \num{5e2} & \num{-1e2} & \num{1e2} \\
    $T_{\Delta y}$ & $\si{\newton}$ & \num{-5e2} & \num{5e2} & \num{-1e2} & \num{1e2} \\
    $T_{\Delta z}$ & $\si{\newton}$ & \num{-5e2} & \num{5e2} & \num{-1e2} & \num{1e2} \\
    $\delta_\mathrm{ab}$ & $\si{\degree}$ & \num{0} & \num{90} & \num{-30} & \num{30} \\
    \midrule
    \multicolumn{6}{ c }{Reference Model Variables} \\
    \midrule
    $\vec{v}^C_\mathrm{x,d}$ & $\si{\meter\per\second}$ & \num{-5} & \num{55} & \num{-2.5} & \num{5} \\
    $\vec{v}^C_\mathrm{y,d}$ & $\si{\meter\per\second}$ & \( -\frac{1}{M} \) & \( \frac{1}{M} \)  & \( -\frac{1}{M} \) & \( \frac{1}{M} \) \\
    $\vec{v}^C_\mathrm{z,d}$ & $\si{\meter\per\second}$ & \num{-5} & \num{5}  & \num{-5} & \num{5} \\
    $\dot{\vec{v}}^C_\mathrm{x,d}$ & $\si{\meter\per\second}$ & \num{-2.5}   & \num{5}   & \num{-5} & \num{5} \\
    $\dot{\vec{v}}^C_\mathrm{y,d}$ & $\si{\meter\per\second}$ & \( -\frac{1}{M} \) & \( \frac{1}{M} \) & \( -\frac{1}{M} \) & \( \frac{1}{M} \) \\
    $\dot{\vec{v}}^C_\mathrm{z,d}$ & $\si{\meter\per\second}$ & \num{-5} & \num{5} & \num{-5} & \num{5} \\
    $\phi_\mathrm{d}$   & $\si{\degree}$ & \num{-45} & \num{45} & \num{-30} & \num{30} \\
    $\theta_\mathrm{d}$ & $\si{\degree}$ & \num{-15} & \num{15} & \num{-5} & \num{5} \\
    $\psi_\mathrm{d}$   & $\si{\degree}$ & -M & M & \num{-15} & \num{15} \\
    $\dot{\phi}_\mathrm{d}$   & $\si{\degree}$ & \num{-30} & \num{30} & \num{-10} & \num{10} \\
    $\dot{\theta}_\mathrm{d}$ & $\si{\degree}$ & \num{-5} & \num{5} & \num{-10} & \num{10} \\
    $\dot{\psi}_\mathrm{d}$   & $\si{\degree}$ & \num{-15} & \num{15} & \num{-10} & \num{10} \\
    \midrule
    \multicolumn{6}{ c }{Inceptor Variables} \\
    \midrule
    $\delta_\mathrm{x}$ & $\si{1}$ & \num{-1} (\num{-2}) & \num{1} (\num{2}) & \num{-10} & \num{10} \\
    $\delta_\mathrm{y}$ & $\si{1}$ & \num{-1} (\num{-2}) & \num{1} (\num{2}) & \num{-10} & \num{10} \\
    $\delta_\mathrm{z}$ & $\si{1}$ & \num{-1} (\num{-2}) & \num{1} (\num{2}) & \num{-10} & \num{10} \\
    $\delta_\mathrm{h}$ & $\si{1}$ & \num{-1} (\num{-2}) & \num{1} (\num{2}) & \num{-10} & \num{10} \\ 
    \midrule
    \multicolumn{6}{ c }{Time Variable} \\
    \midrule
    $t_f$ & $\si{\second}$ & \num{1} & \num{300} &   &    \\
    \bottomrule
\end{tabular}
\caption{State, input, and time variables of the optimal control problem with boundaries and rate constraints (RC). Inceptor limits are doubled for the force-feedback case. \( M \) denotes a large value as replacement for infinity.}
\label{tab:Vars}
\end{table}
\section*{Statements and Declarations}
On behalf of all authors, the corresponding author states that there is no conflict of interest.
\paragraph*{Funding}
The authors declare that no funds, grants, or other support were received during the preparation of this manuscript.
\paragraph*{Competing Interests}
The authors have no relevant financial or non-financial interests to disclose.
\paragraph*{Author Contribution}
All authors took part in planning and designing the study. Daniel Milz led the research, implementation, data collection, analysis, and wrote the manuscript. Marc May contributed to the flight dynamics and optimal control framework, helped write those sections, and discussed the methods and results. Andreas Seefried and Tobias Bellmann supervised the project, provided access to the motion simulator, and helped write the relevant sections. All authors read, reviewed, and approved the final manuscript. 
\paragraph*{Data Availability}
The main part of this study introduces a new approach, explained in the article, that does not need extra data. All data and methods necessary to replicate the findings demonstrating the proposed approach are included in this article or cited.
\end{appendices}

\end{document}